\documentstyle[aps,prb,epsf,floats,twocolumn]{revtex}

\begin{document}
\draft

\twocolumn[\hsize\textwidth\columnwidth\hsize\csname @twocolumnfalse\endcsname

\title{The role of winding numbers in quantum Monte Carlo simulations}
\author{Patrik Henelius and S. M. Girvin}
\address{Department of Physics,
Indiana University, Bloomington, Indiana 47405}
\author{Anders W. Sandvik}
\address{Department of Physics, University of Illinois at Urbana-Champaign,
Urbana, Illinois 61801}
\date{\today}
\maketitle

\begin{abstract}
We discuss the effects of fixing the winding number in quantum Monte Carlo 
simulations. We present a simple geometrical argument as well as strong 
numerical evidence that one can obtain exact ground state results for 
periodic boundary conditions without changing the winding number.
However, for very small systems the temperature has to be considerably lower 
than in simulations with fluctuating winding numbers. The relative deviation 
of a calculated observable from the exact ground state result typically scales 
as $T^{\gamma}$, where the exponent $\gamma$ is model and observable
dependent and the prefactor decreases with increasing system size. 
Analytic results for a quantum rotor model further support our claim.
\end{abstract}
\vskip2mm]

One approach to numerical many-body physics is to stochastically sample 
the ``world line'' configurations of a real-space Euclidean path integral. 
The most common of these quantum Monte Carlo (QMC) methods 
\cite{reviews,ceperley} are based on the Trotter decomposition formula 
in discretized imaginary time,\cite{trotter} 
with a resulting systematic error that vanishes 
as the discretization $\Delta\tau$ is decreased.\cite{suzuki,hirsch} For 
lattice models, methods have recently been constructed for simulations in 
continuous imaginary time,\cite{beard,prokofev,sandvik1} hence directly giving
results exact to within statistical fluctuations. A related method is the 
stochastic series expansion technique \cite{sandvik2} (a generalization of 
Handscomb's method \cite{handscomb}), which samples the power series expansion
of the density matrix, and also is exact. For all these methods, the 
configurations for periodic systems can be labeled by a topological ``winding 
number'' $w$, which counts the net number of times the world lines wrap
around the system in the course of propagation in imaginary time.
In practice, it is often not possible to sample all winding number 
sectors, since changing $w$ requires simultaneous modification of a number 
$\sim L$ world lines, where $L$ is the linear size of the system, with
a resulting low acceptance rate if $L$ is large. Such
non-ergodicity is clearly related to the boundary condition\cite{hirsch,marcu}
(restricting to, e.g., $w=0$ can be considered a particular boundary 
condition), and therefore 
results scaled to infinite system size are still correct, although there can 
be significant deviations from the exact periodic boundary condition results 
for any given small system size. The winding number is a consequence of the 
path integral formulation of quantum mechanics, and a fixed $w$ is not  
related to the Hamiltonian in a simple way. In many cases it is of interest 
to obtain approximation-free results for periodic boundary conditions, 
specifically, and the restriction to fixed $w$ can then be a practical 
limitation of the QMC method.

We here point out that the exact ground state can in fact be 
obtained in any winding number sector, even for the smallest possible systems.
We base our claim on simple geometric considerations, and provide supporting 
simulation results for several many-body Hamiltonians (one- and two 
dimensional Heisenberg spin models) as well as analytic results for a 
quantum rotor. We expect our arguments to be generally valid for models
for which the path integral is positive definite (i.e., there are no sign
problems \cite{hirsch,loh}), which is the case for, e.g., 1D fermion 
systems and non-frustrated spin and boson systems in any dimension.

In order to define the winding number, we first consider the standard path 
integral formulation of a continuous 1D system. The partition 
function for a single particle of mass $M$ in a potential $V(x)$ is
\cite{feynman}
\begin{equation}
Z=\int_{x(\beta )=x(0)} D[x(\tau)]e^{-S} ,
\label{pathz}
\end{equation}
where the action is
\begin{equation}
S=\int_{0}^{\beta } d\tau \left [\frac{M}{2} 
\left ( \frac{dx(\tau)}{d\tau} \right )^2 + V(x(\tau) ) \right ] .
\end{equation}
The integral in (\ref{pathz}) is over all 
paths (or world lines) starting at position $x$ at imaginary time $\tau=0$ and 
ending at the same position $x$ at $\tau=\beta$, where $\beta$ denotes the 
inverse temperature. For a periodic system, the paths can be divided into 
topologically distinct classes, characterized by a winding number $w$ defined
as the net number of times the world line spatially wraps around the 
system. For a many-particle system, the winding number
is defined as the net total displacement to the ``left'' or the ``right'' of 
the world lines in the course of propagation between $\tau=0$ and 
$\tau=\beta$, divided by the length of the system. In higher dimensions, 
there is a winding number associated with each dimension, and the definition 
of these is a direct generalization of the above definition in one dimension.

For interacting many-body systems, analogous real-space path-integrals 
(or related sums \cite{sandvik2} based on series expanding the density matrix 
operator ${\rm e}^{-\beta H}$) can be constructed, and are suitable 
for numerical simulations in cases where the weight associated with the paths
is positive definite. Such QMC methods have been developed, e.g., for 
lattice fermions in one dimension \cite{hirsch}, lattice\cite{batr} and 
continuum\cite{ceperley} bosons, and quantum spin systems.\cite{suzuki} 
In the case of fermions
in higher dimensions, the path integral cannot be efficiently sampled 
directly, due to the non-positive definiteness of the weight, which leads 
to the infamous ``sign problem''.\cite{loh,blankenbecler}

Stochastic sampling of the bosonic paths within a sector of a fixed winding
number can be accomplished by local modifications of the world lines, and is 
typically a relatively straight-forward task. The global modifications 
required in order to change the winding number are often practically 
impossible to carry out efficiently, however. Recently ``loop-cluster''
algorithms have been developed which in principle automatically sample 
all winding number sectors.\cite{evertz} However, such algorithms do not 
perform well for all models, and therefore the restriction to a sector with 
fixed winding number remains the only option in many cases. 
We note that the winding number itself is related to long-range coherence in 
the system. For boson and spin systems, the winding number fluctuation 
$\langle w^2\rangle$ is directly proportional to the superfluid density and 
the spin stiffness, respectively.\cite{ceperley}
In some cases these quantities can, however, also be computed in a restricted 
simulation. \cite{batr,ceperley}

We here argue that the exact ground state can 
actually be studied in any $w$ sector. 
Consider first again a 1D periodic system with a single 
particle. A path integral configuration can be visualized as a string on the 
surface of a cylinder with periodic boundary conditions (i.e., a torus). As 
the temperature is lowered the length $\beta =1/T$ of the system in the 
imaginary time direction becomes much larger than the spatial system size, 
and the string can then wrap around the cylinder multiple times in both 
directions between $\tau=0$ and $\beta$. In the winding number $w$ sector, 
the net number of revolutions has to be $w$. Nevertheless, locally, in an 
imaginary-time segment of length $\Delta\tau \ll \beta$, it would not be 
possible to detect the 
effects of this restriction. Hence, any quantity that can be defined on a 
finite segment of the cylinder should be the same in any winding number sector
as $\beta \to \infty$. Since correlation functions decay exponentially with 
the imaginary-time separation in a finite system, due to the finite-size gap, 
all quantities which are not defined in terms of the global winding number 
itself should become exact as $\beta \to \infty$. This argument can clearly 
be generalized for a many-body system in any dimension, again of course 
provided that all paths have positive weights. In the case of mixed signs 
the positive and negative contributions to the partition function cancel 
as $\beta \to \infty$, and physical quantities are given by finite ratios 
of two vanishing numbers. It is then not clear that all the winding number 
sectors become equal as $\beta \to \infty$ (although we have not proved
the contrary), and we will not consider this  intricate issue further here.

We can rigorously prove that fixed $w$ gives the correct ground state
of a single particle in one dimension. 
This system (with $ML^2=2\pi^2$) is equivalent 
to the quantum rotor, described by the Hamiltonian
\begin{equation}
H=-\partial^2/\partial\theta^2 .
\end{equation}
The eigenfunctions labeled by
the angular momentum $m$ are
\begin{equation}
\Psi_m (\theta) =\exp(im\theta).
\end{equation}
The partition function is 
\begin{equation}
Z=\sum_{m=-\infty}^{+\infty}\exp(-\beta m^2),
\label{partm}
\end{equation}
and can be transformed from a sum over angular momenta to a sum over winding 
numbers in the following way: In the discrete path integral formulation 
with $\Delta\tau = \beta/N$ the partition function is 
\begin{eqnarray}
Z & = & \int D\theta\prod_{j=1}^N\langle\theta_{j+1}\vert
\exp(-\Delta\tau H)\vert\theta_{j}\rangle \nonumber \\
& = & \int D\theta\prod_{j=1}^N \left [ \sum_m\exp(-\Delta\tau m^2)
\exp[im(\theta_{j+1}-\theta_{j})] \right ],
\end{eqnarray}
where $\Delta\tau=\beta/N$ and $\theta_{N+1}=\theta_1$. 
Using Poisson's summation formula,
\begin{equation}
\sum_{m=-\infty}^{\infty}f(m)=\sum_{n=-\infty}^{\infty}F(2\pi n),
\label{poisson}
\end{equation}
where $F(n)$ is the Fourier transform of $f(m)$, we obtain 
\begin{equation}
Z=\int D\theta\prod_{j=1}^N  \left [  \sum_n 
\exp \left [- {(\theta_{j+1}-\theta_{j}-2\pi n)^2 \over 4\Delta\tau} 
\right ] \right ].
\label{afterp}
\end{equation}
Next we write the $j$ (or $\tau$) dependence of $\theta$ as
\begin{equation}
\theta_j=\bmod[\theta_1+{2\pi w\over N}(j-1) +\delta\theta_j, 2\pi],
\end{equation}
where $w$ is the winding number. For the time being we neglect the 
fluctuations $\delta\theta_j$, which are the same for all $w$. 
As $\Delta\tau \to 0$, only the term $n=0$ contributes in the
above sum, unless $\theta$ crosses the boundary at $\theta=0$,
in which case $n=-1$ or $n=+1$ will give the contributing term. 
Hence the partition function can be expressed as
\begin{equation}
Z=f(\beta) \sum_{w=-\infty}^{\infty}\exp(-\pi^2w^2/ \beta),
\end{equation}
where $f(\beta)$ is a function containing the effects of the fluctuations 
$\delta\theta_j$ which we have so far neglected. The easiest way to find
$f(\beta)$ is to simply equate the above result with Eq. (\ref{partm}).
This gives our final result for the partition function expressed
as a sum over winding numbers:
\begin{equation}
Z=\sum_{w=-\infty}^{\infty}\sqrt{\pi/\beta}\exp(-{\pi^2w^2/ \beta}).
\label{partw}
\end{equation}
This result could have been directly obtained using Poisson's summation 
formula, Eq. (\ref{poisson}), on Eq.~(\ref{partm}). The above derivation
shows that the new summation index indeed is the winding number, which
hence can be thought of as a quantity conjugate to the angular momentum.

\begin{figure}[ht]
\centering
\epsfxsize=8cm
\leavevmode
\epsffile{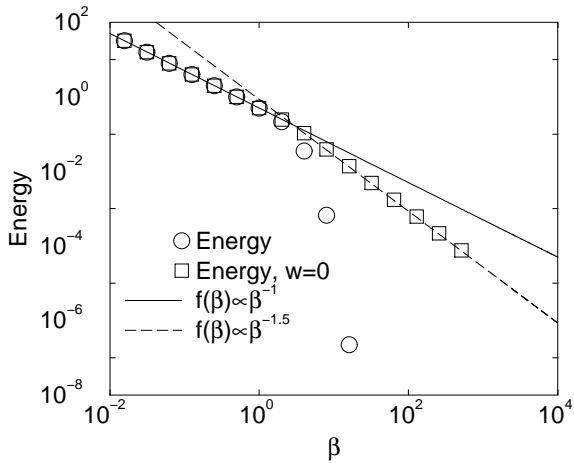}
\caption{Rotor model : The full energy and the energy
for the $w=0$ sector vs. $\beta$. The lines
are the assymptotic high and low (for $w=0$) temperature forms.}
\label{fig01}
\end{figure}

The energy  $E=-(1/Z) \partial Z /\partial \beta$ given  as a sum over 
angular momenta is
\begin{equation}
E=-{1\over Z} \sum_m m^2\exp(-\beta m^2),
\end{equation}
and expressed as a sum over winding numbers 
\begin{equation}
E=-{1\over Z} \sum_w \sqrt{\pi} ({1\over 2\beta^{3/2}}-{\pi^2w^2\over 
\beta^{5/2}})
{\rm e}^{-\pi^2w^2/\beta}.
\label{energyw}
\end{equation}
In Fig. \ref{fig01} the energy of the rotor is plotted versus $\beta$
on a log-log scale. It decreases as $\beta^{-1}$ at high temperatures,
changing to an exponential decay around $\beta=1$. In the same graph
we also show the energy evaluated in the $w=0$ sector. 
At high temperatures the results coincide with the full energy, but 
around $\beta=1$ the behavior changes to be of the form $\beta^{-3/2}$, 
instead of exponential, as can be readily extracted from Eq.~(\ref{energyw}).
Hence, we indeed get the correct energy (namely, zero) in the $w=0$ sector,
but the approach to the ground state is much slower than in the
``ensemble'' with fluctuating $w$.

Next we present numerical results for several many-body models. We study 
the Heisenberg model, defined by the Hamiltonian
\begin{equation}
H= \sum_{\langle i,j\rangle} {\bf S}_i \cdot {\bf S}_{j},
\end{equation}
where $\langle i,j\rangle$ indicates a pair of nearest neighbors lattice
sites, and ${\bf S}_i$ is a spin-$S$ operator at site $i$. We consider the 
1D model with $S=1/2$ and $S=1$, as well as the 2D 
model with $S=1/2$. We note  that these quantum spin models are formally
equivalent to constrained boson systems. 

For the simulations we use the stochastic series expansion algorithm,
\cite{sandvik2} which is based on a power series expansion of 
e$^{-\beta H}$, and hence is not a standard path integral method. 
The configuration space is nevertheless very strongly related 
to an Euclidean path integral,\cite{sandvik1} and the winding number has 
exactly the same meaning. For systems of linear size $L \alt 16$ Monte Carlo 
updates changing the winding number can be easily carried out, but for larger
systems a restriction to fixed $w$ is necessary in practice. The advantage
of the method is that there are no other sources of systematic errors. We 
present energy results both for $w=0$ and fluctuating $w$, and compare with 
exact diagonalization data. 

The simulation scheme is formulated in the standard basis where the operators
$S^z_i$ are diagonal. The internal energy per spin can be 
calculated in two different ways in the simulation; from the nearest-neighbor 
correlation function $\langle S^z_{i}S^z_{i+1}\rangle$ as well as from a 
manifestly rotationally invariant estimator giving the full 
$\langle {\bf S}_{i} \cdot {\bf S}_{i+1}\rangle$.\cite{sandvik2}
We define
\begin{mathletters}
\begin{eqnarray}
E_z & = & 3 \langle S^z_{i}S^z_{i+1}\rangle, \\
E_s & = &  \langle {\bf S}_{i} \cdot {\bf S}_{i+1}\rangle.
\end{eqnarray}
\end{mathletters}
The agreement between the two estimates can serve as a good internal check of 
the spherical spin symmetry in the simulation. With $w$ fixed in a finite
system, the spin-rotational symmetry  is broken since only the $xy$-terms are 
involved in the spatial propagation (spin flipping) of the path
and the estimates will therefore not agree completely. 

\begin{figure}[ht]
\centering
\epsfxsize=8cm
\leavevmode
\epsffile{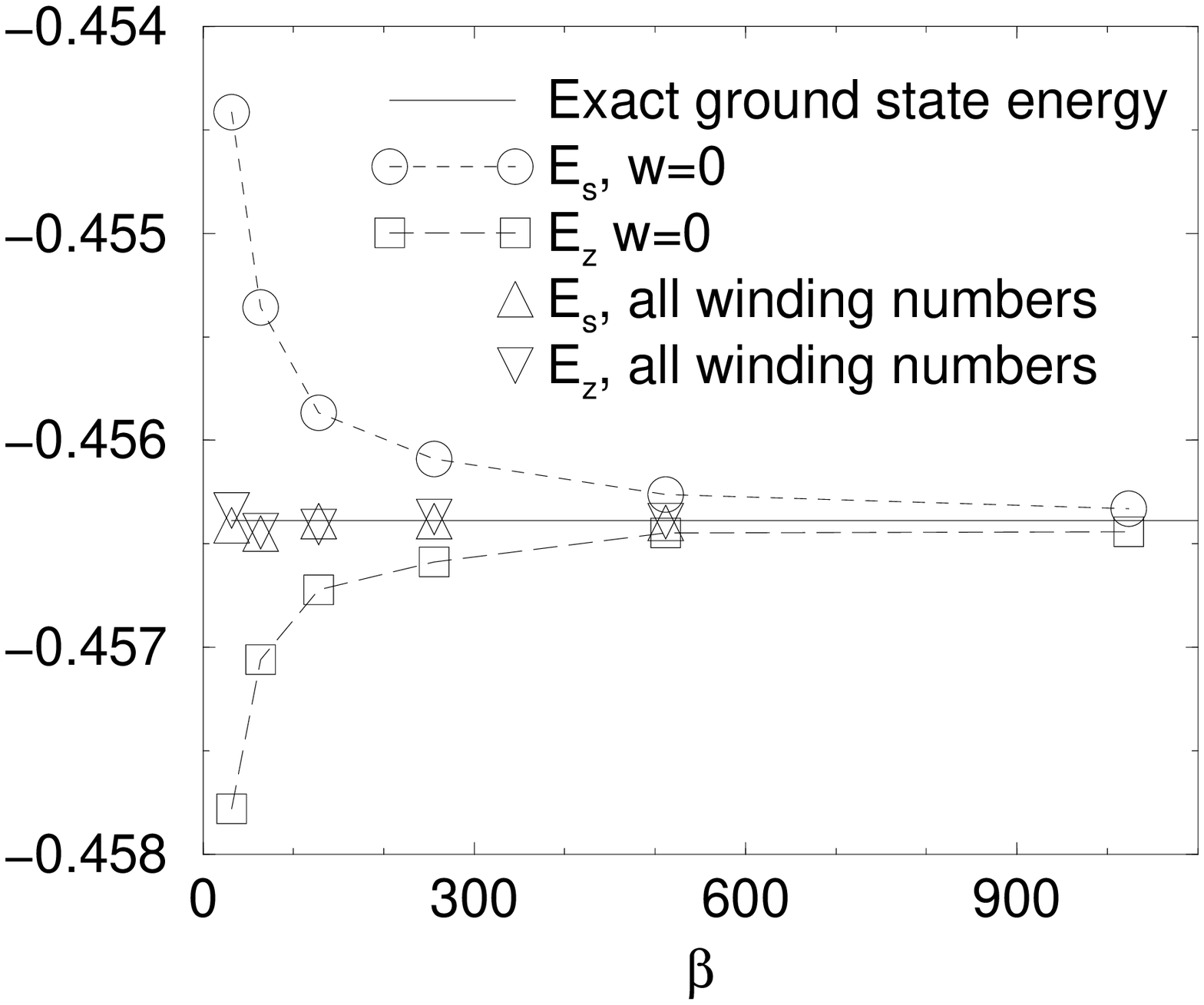}
\caption{1D spin-1/2 model: The energy estimators $E_s$ and $E_z$ 
vs. $\beta$ for L=8.}
\label{fig02}
\end{figure}

In Fig.~\ref{fig02}  $E_s$ and $E_z$
are graphed versus the inverse temperature $\beta$ used 
in simulations of an 8-site $S=1/2$ chain, both in the $w=0$ sector and with 
fluctuating winding numbers. In the $w=0$ sector, the rotationally invariant 
and diagonal estimates approach the exact ground state result from above, 
and below, respectively. For fluctuating winding numbers the exact 
ground state energy is obtained within statistical errors already for 
$\beta \approx 8$ for this system size, reflecting the large finite-size gap 
and the exponential approach to the ground state with increasing $\beta$. 

\begin{figure}[ht]
\centering
\epsfxsize=8cm
\leavevmode
\epsffile{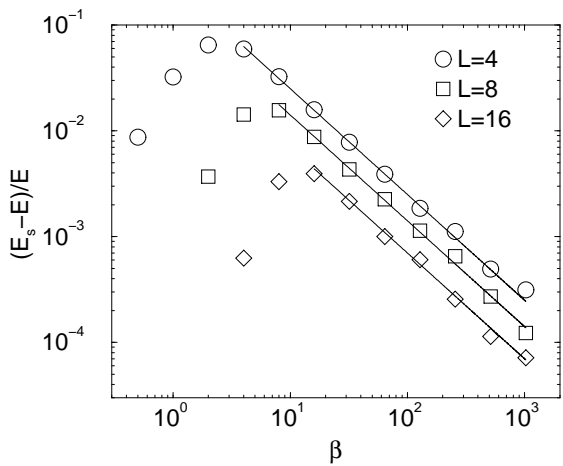}
\caption{1D spin-1/2 model: The deviation from the 
exact energy of $E_s$ vs. $\beta$ for L=4,8 and 16.
The lines have slope -1.}
\label{fig03}
\end{figure}

In Fig. \ref{fig03} the difference between Monte Carlo data obtained in the 
$w=0$ sector and the exact temperature-dependent energy is graphed on a 
log-log scale for $L=4,8$ and 16. At high temperatures the difference
tends to vanish rapidly since the system is too short in the imaginary 
time dimension to allow for the winding number to change from zero. At low 
temperatures the deviation decreases as $\beta^{-\gamma}$, with $\gamma=1$
within the numerical precision, and there is a maximum at an
intermediate value $\beta^{\ast}$. As expected, the maximum
difference decreases rapidly with system size. The relative error
decreases roughly as $1/L$ at a given temperature, and $\beta^{\ast}$ 
increases roughly as $L$. $\beta^{\ast}$ hence reflects the inverse 
finite-size gap. These results confirm that for a given desired accuracy of 
the energy at a given temperature, there will be some system size beyond which
it is not necessary to sample different winding number sectors. Furthermore, 
the exact ground state is always obtained as $\beta \to \infty$. However,
for small systems we have to use higher values of $\beta$ in order to 
achieve a certain accuracy if we do not sample all winding number sectors. 

\begin{figure}[ht]
\centering
\epsfxsize=8cm
\leavevmode
\epsffile{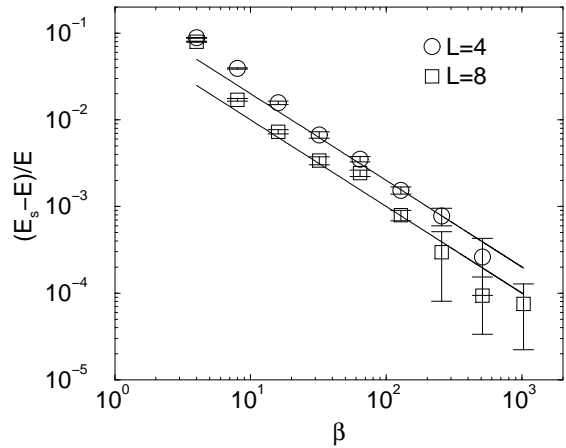}
\caption{1D spin-1 model: The deviation from the 
ground state energy of $E_s$ vs. $\beta$ 
for L=4 and 8. The lines have slope -1.}
\label{fig04}
\end{figure}

\begin{figure}[ht]
\centering
\epsfxsize=8cm
\leavevmode
\epsffile{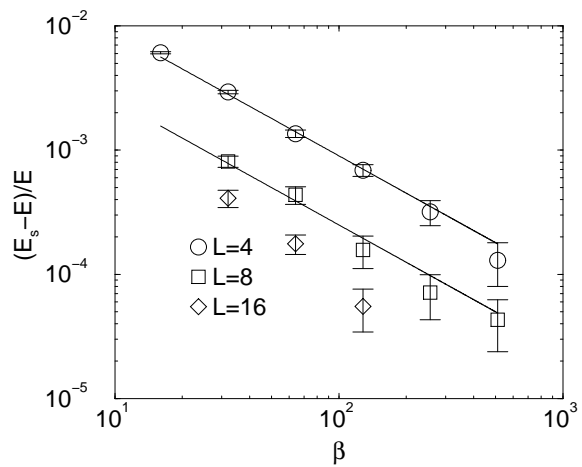}
\caption{2D spin-1/2 model: The deviation from the 
ground state energy of $E_s$ vs. $\beta$ 
for L=4,8 and 16. The lines have slope -1}
\label{fig05}
\end{figure}

Next we consider the 1D $S=1$ model and the 2D 
$S=1/2$ model. The relative deviation of the $w=0$ energies from the exact 
ground state values are graphed versus $\beta$ for different system sizes in
Fig. \ref{fig04} and Fig. \ref{fig05} (for the 2D systems
with $L > 4$ ``exact'' results were obtained in simulations with 
fluctuating $w$). Again we see that at low temperatures the ground state is 
approached as $\beta^{-1}$, instead of the exponential approach expected 
for the exact energy.

To summarize we have given an intuitive argument why restricting the winding 
number in QMC simulations number should not affect calculated ground state 
properties of even the smallest lattices. We have given strong numerical 
evidence to 
support this statement, in addition to rigorous results for a simple quantum
rotor model. Typically, the asymptotic deviations from the exact periodic 
boundary condition results scale as $\beta^{-\gamma}$ at low temperatures, 
with a prefactor that goes to zero as the system size increases. Hence, 
in terms of the $\beta$ needed to obtain the ground state, to the statistical
precision possible in QMC simulations, there does not appear to be any 
advantage of fluctuating numbers for moderate and large systems. 

Our considerations are closely related to the imaginary-time boundary 
conditions recently discussed by T\"auber and Nelson.\cite{tauber} They showed 
that the condition that the world line configurations at $\tau =0$ and 
$\tau=\beta$ being equal can be relaxed, still giving the same result as the 
periodic imaginary time boundary condition when $\beta \to \infty$. In 
analogy with our discussion above, the reason for this is that the changed 
boundary condition does not affect the behavior of the world lines in a 
segment of length $\ll \beta$.

This work was supported by the National Science Foundation under Grants
No. DMR 97-14055, CDA-9601632 and DMR-97-12765. P.H. acknowledges
support by the Ella och Georg Ehrnrooths stiftelse.

\end{document}